
%
%
%
%
%
%
%
\def\standardrisposta{s }\def\reducedrisposta{r }
\def\doublerisposta{d }\def\cartarisposta{e }\def\amsrisposta{y }
\newcount\ingrandimento \newcount\sinnota \newcount\dimnota
\newcount\unoduecol \newdimen\collhsize \newdimen\tothsize
\newdimen\fullhsize \newcount\controllorisposta
\newskip\infralinea  \global\controllorisposta=0
\message{ ********    Welcome to PANDA macros (Plain TeX, AP, 1991)}
\message{ ******** }
\message{       You'll have to answer a few questions in lowercase.}
\message{>  Do you want it in double-page (d), reduced (r)}
\message{or standard format (s) ? }\read-1 to\risposta
\message{>  Do you want it in USA A4 (u) or EUROPEAN A4 (e)}
\message{paper size ? }\read-1 to\srisposta
\message{>  Do you have AMS (math) fonts (y/n) ? }\read-1 to\arisposta
%
%
%
%
%
\ifx\risposta\standardrisposta
\message{>> This will come out UNREDUCED << }
\ingrandimento=1200 \dimnota=2 \unoduecol=1 \infralinea=16pt
\global\controllorisposta=1 \fi
\ifx\risposta\reducedrisposta
\ingrandimento=1095 \dimnota=1 \unoduecol=1 \infralinea=14pt
\message{>> This will come out REDUCED << }
\global\controllorisposta=1 \fi
\ifx\risposta\doublerisposta
\ingrandimento=1000 \dimnota=2 \unoduecol=2 \infralinea=12pt
\message{>> You must print this in LANDSCAPE orientation << }
\global\controllorisposta=1 \fi
\ifnum\controllorisposta=0
\message{>>> ERROR IN INPUT, I ASSUME STANDARD UNREDUCED FORMAT <<< }
\ingrandimento=1200 \dimnota=2 \unoduecol=1 \infralinea=16pt\fi
\def\magnificazione{\magnification=\ingrandimento}  \magnificazione
\ifx\risposta\doublerisposta
\ifx\srisposta\cartarisposta
\tothsize=25.5truecm \collhsize=12.0truecm \vsize=17.0truecm \else
\tothsize=9.4truein \collhsize=4.4truein \vsize=6.8truein \fi \else
\ifx\srisposta\cartarisposta
\tothsize=6.5truein \vsize=24truecm \else
\tothsize=6.5truein \vsize=8.9truein \fi
\collhsize=4.4truein \fi
\tolerance=10000
\leftskip=0pt \rightskip=0pt \sinnota=1 \parskip 0pt plus 2pt
%
%
%
%
\font\ninerm=cmr9  \font\eightrm=cmr8  \font\sixrm=cmr6
\font\ninei=cmmi9  \font\eighti=cmmi8  \font\sixi=cmmi6
\font\ninesy=cmsy9  \font\eightsy=cmsy8  \font\sixsy=cmsy6
\font\ninebf=cmbx9  \font\eightbf=cmbx8  \font\sixbf=cmbx6
\font\ninett=cmtt9  \font\eighttt=cmtt8 \font\nineit=cmti9
\font\eightit=cmti8 \font\ninesl=cmsl9  \font\eightsl=cmsl8
\skewchar\ninei='177 \skewchar\eighti='177 \skewchar\sixi='177
\skewchar\ninesy='60 \skewchar\eightsy='60 \skewchar\sixsy='60
\hyphenchar\ninett=-1 \hyphenchar\eighttt=-1 \hyphenchar\tentt=-1
%
\font\tencaps=cmcsc10              
%
\ifx\arisposta\amsrisposta
\font\mmmath=msym10                
\else \gdef\mmmath{\bf}  \fi
\ifnum\ingrandimento=1095

\font\capsone=cmcsc10 at 10.95pt 

\else

\font\capsone=cmcsc10 at 12pt 
\fi
\def\ttaarr{\bf}		
\def\ppaarr{\sl}		
\catcode`@=11
\newskip\ttglue
\gdef\tenpoint{\def\rm{\fam0\tenrm}
  \textfont0=\tenrm \scriptfont0=\sevenrm \scriptscriptfont0=\fiverm
  \textfont1=\teni \scriptfont1=\seveni \scriptscriptfont1=\fivei
  \textfont2=\tensy \scriptfont2=\sevensy \scriptscriptfont2=\fivesy
  \textfont3=\tenex \scriptfont3=\tenex \scriptscriptfont3=\tenex
  \textfont\itfam=\tenit \def\it{\fam\itfam\tenit}
  \textfont\slfam=\tensl \def\sl{\fam\slfam\tensl}
  \textfont\ttfam=\tentt \def\tt{\fam\ttfam\tentt}
  \textfont\bffam=\tenbf \scriptfont\bffam=\sevenbf
  \scriptscriptfont\bffam=\fivebf \def\bf{\fam\bffam\tenbf}
  \tt \ttglue=.5em plus.25em minus.15em
  \normalbaselineskip=12pt
  \setbox\strutbox=\hbox{\vrule height8.5pt depth3.5pt width0pt}
  \let\sc=\eightrm \let\big=\tenbig \normalbaselines\rm}
\gdef\ninepoint{\def\rm{\fam0\ninerm}
  \textfont0=\ninerm \scriptfont0=\sixrm \scriptscriptfont0=\fiverm
  \textfont1=\ninei \scriptfont1=\sixi \scriptscriptfont1=\fivei
  \textfont2=\ninesy \scriptfont2=\sixsy \scriptscriptfont2=\fivesy
  \textfont3=\tenex \scriptfont3=\tenex \scriptscriptfont3=\tenex
  \textfont\itfam=\nineit \def\it{\fam\itfam\nineit}
  \textfont\slfam=\ninesl \def\sl{\fam\slfam\ninesl}
  \textfont\ttfam=\ninett \def\tt{\fam\ttfam\ninett}
  \textfont\bffam=\ninebf \scriptfont\bffam=\sixbf
  \scriptscriptfont\bffam=\fivebf \def\bf{\fam\bffam\ninebf}
  \tt \ttglue=.5em plus.25em minus.15em
  \normalbaselineskip=11pt
  \setbox\strutbox=\hbox{\vrule height8pt depth3pt width0pt}
  \let\sc=\sevenrm \let\big=\ninebig \normalbaselines\rm}
\gdef\eightpoint{\def\rm{\fam0\eightrm}
  \textfont0=\eightrm \scriptfont0=\sixrm \scriptscriptfont0=\fiverm
  \textfont1=\eighti \scriptfont1=\sixi \scriptscriptfont1=\fivei
  \textfont2=\eightsy \scriptfont2=\sixsy \scriptscriptfont2=\fivesy
  \textfont3=\tenex \scriptfont3=\tenex \scriptscriptfont3=\tenex
  \textfont\itfam=\eightit \def\it{\fam\itfam\eightit}
  \textfont\slfam=\eightsl \def\sl{\fam\slfam\eightsl}
  \textfont\ttfam=\eighttt \def\tt{\fam\ttfam\eighttt}
  \textfont\bffam=\eightbf \scriptfont\bffam=\sixbf
  \scriptscriptfont\bffam=\fivebf \def\bf{\fam\bffam\eightbf}
  \tt \ttglue=.5em plus.25em minus.15em
  \normalbaselineskip=9pt
  \setbox\strutbox=\hbox{\vrule height7pt depth2pt width0pt}
  \let\sc=\sixrm \let\big=\eightbig \normalbaselines\rm}
\gdef\tenbig#1{{\hbox{$\left#1\vbox to8.5pt{}\right.\n@space$}}}
\gdef\ninebig#1{{\hbox{$\textfont0=\tenrm\textfont2=\tensy
   \left#1\vbox to7.25pt{}\right.\n@space$}}}
\gdef\eightbig#1{{\hbox{$\textfont0=\ninerm\textfont2=\ninesy
   \left#1\vbox to6.5pt{}\right.\n@space$}}}
%
\newbox\fotlinebb \newbox\hedlinebb \newbox\leftcolumn
\gdef\makeheadline{\vbox to 0pt{\vskip-22.5pt
     \fullline{\vbox to8.5pt{}\the\headline}\vss}\nointerlineskip}
\gdef\makehedlinebb{\vbox to 0pt{\vskip-22.5pt
     \fullline{\vbox to8.5pt{}\copy\hedlinebb\hfil
     \line{\hfill\the\headline\hfill}}\vss}
     \nointerlineskip}
\gdef\makefootline{\baselineskip=24pt \fullline{\the\footline}}
\gdef\makefotlinebb{\baselineskip=24pt
    \fullline{\copy\fotlinebb\hfil\line{\hfill\the\footline\hfill}}}
\gdef\doubleformat{\shipout\vbox{\makehedlinebb
     \fullline{\box\leftcolumn\hfil\columnbox}\makefotlinebb}
     \advancepageno}

\gdef\columnbox{\leftline{\pagebody}}
\gdef\line#1{\hbox to\hsize{\hskip\leftskip#1\hskip\rightskip}}
\gdef\fullline#1{\hbox to\fullhsize{\hskip\leftskip{#1}%
\hskip\rightskip}}
\gdef\footnote#1{\let\@sf=\empty
         \ifhmode\edef\#sf{\spacefactor=\the\spacefactor}\/\fi
         #1\@sf\vfootnote{#1}}
\gdef\vfootnote#1{\insert\footins\bgroup
         \ifnum\dimnota=1  \eightpoint\fi
         \ifnum\dimnota=2  \ninepoint\fi
         \ifnum\dimnota=0  \tenpoint\fi
         \interlinepenalty=\interfootnotelinepenalty
         \splittopskip=\ht\strutbox
         \splitmaxdepth=\dp\strutbox \floatingpenalty=20000
         \leftskip=\oldssposta \rightskip=\olddsposta
         \spaceskip=0pt \xspaceskip=0pt
         \ifnum\sinnota=0   \textindent{#1}\fi
         \ifnum\sinnota=1   \item{#1}\fi
         \footstrut\futurelet\next\fo@t}
\gdef\fo@t{\ifcat\bgroup\noexpand\next \let\next\f@@t
             \else\let\next\f@t\fi \next}
\gdef\f@@t{\bgroup\aftergroup\@foot\let\next}
\gdef\f@t#1{#1\@foot}
\gdef\@foot{\strut\egroup}
\gdef\footstrut{\vbox to\splittopskip{}}
\skip\footins=\bigskipamount
\count\footins=1000  \dimen\footins=8in
\catcode`@=12
\tenpoint  \baselineskip=\infralinea
\newskip\olddsposta \newskip\oldssposta
\global\oldssposta=\leftskip \global\olddsposta=\rightskip

\gdef\yespagenumbers{\footline={\hss\tenrm\folio\hss}}
\gdef\ciao{\par\vfill\supereject
      \ifnum\unoduecol=2 \if R\lrcol \null\vfill\eject \fi\fi \end}

\ifnum\unoduecol=1 \hsize=\tothsize   \fullhsize=\tothsize \fi
\ifnum\unoduecol=2 \hsize=\collhsize  \fullhsize=\tothsize \fi
\global\let\lrcol=L
\ifnum\unoduecol=1 \output{\plainoutput}\fi
\ifnum\unoduecol=2 \output{\if L\lrcol
       \global\setbox\leftcolumn=\columnbox
       \global\setbox\fotlinebb=\line{\hfill\the\footline\hfill}
       \global\setbox\hedlinebb=\line{\hfill\the\headline\hfill}
       \advancepageno
      \global\let\lrcol=R \else \doubleformat \global\let\lrcol=L \fi
       \ifnum\outputpenalty>-20000 \else\dosupereject\fi}\fi
\def\ifdoublepage{\ifnum\unoduecol=2 }
\def\filldots{\leaders\hbox to 1em{\hss.\hss}\hfill}
\def\inquadrb#1 {\vbox {\hrule  \hbox{\vrule \vbox {\vskip .2cm
    \hbox {\ #1\ } \vskip .2cm } \vrule  }  \hrule} }

\def\newline{\hfil\break}
\def\jump{\vskip\baselineskip} \newskip\iinnffrr
\def\sjump{\iinnffrr=\baselineskip
          \divide\iinnffrr by 2 \vskip\iinnffrr}
\def\bjump{\vskip\baselineskip \vskip\baselineskip}
\newcount\nmbnota  \def\clearnmbnota{\global\nmbnota=0}
\def\note#1{\global\advance\nmbnota by 1
    \footnote{$^{\the\nmbnota}$}{#1}}  \clearnmbnota
\def\setnote#1{\expandafter\xdef\csname#1\endcsname{\the\nmbnota}}
\newcount\nbmfig  \def\clearnbmfig{\global\nbmfig=0}
\gdef\figure{\global\advance\nbmfig by 1
      {\rm fig. \the\nbmfig}}   \clearnbmfig
\def\setfig#1{\expandafter\xdef\csname#1\endcsname{fig. \the\nbmfig}}
 \def\endformula{\eqno\numero $$}
 \def\efr{\endformula}
\newcount\frmcount \def\clearfrmcount{\global\frmcount=0}
\def\numero{\global\advance\frmcount by 1   \ifnum\indappcount=0
  {\ifnum\cpcount <1 {\hbox{\rm (\the\frmcount )}}  \else
  {\hbox{\rm (\the\cpcount .\the\frmcount )}} \fi}  \else
  {\hbox{\rm (\applett .\the\frmcount )}} \fi}
\def\nameformula#1{\global\advance\frmcount by 1%
\ifnum\draftnum=0  {\ifnum\indappcount=0%
{\ifnum\cpcount<1\xdef\spzzttrra{(\the\frmcount )}%
\else\xdef\spzzttrra{(\the\cpcount .\the\frmcount )}\fi}%
\else\xdef\spzzttrra{(\applett .\the\frmcount )}\fi}%
\else\xdef\spzzttrra{(#1)}\fi%
\expandafter\xdef\csname#1\endcsname{\spzzttrra}
\eqno \ifnum\draftnum=0 {\ifnum\indappcount=0
  {\ifnum\cpcount <1 {\hbox{\rm (\the\frmcount )}}  \else
  {\hbox{\rm (\the\cpcount .\the\frmcount )}}\fi}   \else
  {\hbox{\rm (\applett .\the\frmcount )}} \fi} \else (#1) \fi $$}
\def\nfr{\nameformula}    
\def\nameali#1{\global\advance\frmcount by 1%
\ifnum\draftnum=0  {\ifnum\indappcount=0%
{\ifnum\cpcount<1\xdef\spzzttrra{(\the\frmcount )}%
\else\xdef\spzzttrra{(\the\cpcount .\the\frmcount )}\fi}%
\else\xdef\spzzttrra{(\applett .\the\frmcount )}\fi}%
\else\xdef\spzzttrra{(#1)}\fi%
\expandafter\xdef\csname#1\endcsname{\spzzttrra}
\eqno \ifnum\draftnum=0  {\ifnum\indappcount=0
  {\ifnum\cpcount <1 {\hbox{\rm (\the\frmcount )}}  \else
  {\hbox{\rm (\the\cpcount .\the\frmcount )}}\fi}   \else
  {\hbox{\rm (\applett .\the\frmcount )}} \fi} \else (#1) \fi}
\clearfrmcount
\newcount\cpcount \def\clearcpcount{\global\cpcount=0}
\newcount\subcpcount \def\clearsubcpcount{\global\subcpcount=0}
\newcount\appcount \def\clearappcount{\global\appcount=0}
\newcount\indappcount \def\clearindappcount{\indappcount=0}
\newcount\sottoparcount 

\def\applett{\ifcase\appcount  \or {A}\or {B}\or {C}\or
{D}\or {E}\or {F}\or {G}\or {H}\or {I}\or {J}\or {K}\or {L}\or
{M}\or {N}\or {O}\or {P}\or {Q}\or {R}\or {S}\or {T}\or {U}\or
{V}\or {W}\or {X}\or {Y}\or {Z}\fi
             \ifnum\appcount<0
    \message{>>  ERROR: counter \appcount out of range <<}\fi
             \ifnum\appcount>26
   \message{>>  ERROR: counter \appcount out of range <<}\fi}
\clearappcount  \clearindappcount
\newcount\connttrre  \def\clearconnttrre{\global\connttrre=0}
\newcount\countref  \def\clearcountref{\global\countref=0}
\clearcountref
\def\chapter#1{\global\advance\cpcount by 1 \clearfrmcount
                 \goodbreak\null\jump\nobreak
                 \clearsubcpcount\clearindappcount
                 \itemitem{\ttaarr\the\cpcount .\qquad}{\ttaarr #1}
                 \par\nobreak\jump\sjump\nobreak}
\def\section#1{\global\advance\subcpcount by 1 \goodbreak\null
                  \sjump\nobreak\ifnum\indappcount=0
                 {\ifnum\cpcount=0 {\itemitem{\ppaarr
               .\the\subcpcount\quad\enskip\ }{\ppaarr #1}\par} \else
                 {\itemitem{\ppaarr\the\cpcount .\the\subcpcount\quad
                  \enskip\ }{\ppaarr #1} \par}  \fi}
                \else{\itemitem{\ppaarr\applett .\the\subcpcount\quad
                 \enskip\ }{\ppaarr #1}\par}\fi\nobreak\jump\nobreak}
\clearsubcpcount
\def\appendix#1{\global\advance\appcount by 1 \clearfrmcount
                  \goodbreak\null\jump\nobreak
                  \global\advance\indappcount by 1 \clearsubcpcount
                  \itemitem{\ttaarr App.\applett\ }{\ttaarr #1}
                  \nobreak\jump\sjump\nobreak}
\clearappcount \clearindappcount
\def\references{\goodbreak\null\jump\nobreak
   \itemitem{}{\ttaarr References} \nobreak\jump\sjump\nobreak}
\clearcountref
\def\introduction{\clearindappcount\clearappcount\clearcpcount
                  \clearsubcpcount\goodbreak\null\jump\nobreak
  \itemitem{}{\ttaarr Introduction} \nobreak\jump\sjump\nobreak}

\def\setchap#1{\ifnum\indappcount=0{\ifnum\subcpcount=0%
\xdef\spzzttrra{\the\cpcount}%
\else\xdef\spzzttrra{\the\cpcount .\the\subcpcount}\fi}
\else{\ifnum\subcpcount=0 \xdef\spzzttrra{\applett}%
\else\xdef\spzzttrra{\applett .\the\subcpcount}\fi}\fi
\expandafter\xdef\csname#1\endcsname{\spzzttrra}}
    \newcount\draftnum
\newcount\ppora   \newcount\ppminuti
\global\ppora=\time   \global\ppminuti=\time
\global\divide\ppora by 60  \draftnum=\ppora
\multiply\draftnum by 60    \global\advance\ppminuti by -\draftnum
\global\draftnum=0
\def\droggi{\number\day /\number\month /\number\year\ \the\ppora
:\the\ppminuti}

\global\draftnum=0
%
%
\catcode`@=11
\gdef\Ref#1{\expandafter\ifx\csname @rrxx@#1\endcsname\relax%
{\global\advance\countref by 1%
\ifnum\countref>200%
\message{>>> ERROR: maximum number of references exceeded <<<}%
\expandafter\xdef\csname @rrxx@#1\endcsname{0}\else%
\expandafter\xdef\csname @rrxx@#1\endcsname{\the\countref}\fi}\fi%
\ifnum\draftnum=0 \csname @rrxx@#1\endcsname \else#1\fi}
\gdef\beginref{\ifnum\draftnum=0  \gdef\Rref{\fairef}
\gdef\endref{\scriviref} \else\relax\fi}
\gdef\Rref#1#2{\item{[#1]}{#2}}  \gdef\endref{\relax}
\newcount\conttemp
\gdef\fairef#1#2{\expandafter\ifx\csname @rrxx@#1\endcsname\relax
{\global\conttemp=0
\message{>>> ERROR: reference [#1] not defined <<<} } \else
{\global\conttemp=\csname @rrxx@#1\endcsname } \fi
\global\advance\conttemp by 50
\global\setbox\conttemp=\hbox{#2} }
\gdef\scriviref{\clearconnttrre\conttemp=50
\loop\ifnum\connttrre<\countref \advance\conttemp by 1
\advance\connttrre by 1
\item{[\the\connttrre]}{\unhcopy\conttemp} \repeat}
\clearcountref \clearconnttrre
\catcode`@=12
\def\slashchar#1{\setbox0=\hbox{$#1$} \dimen0=\wd0
     \setbox1=\hbox{/} \dimen1=\wd1 \ifdim\dimen0>\dimen1
      \rlap{\hbox to \dimen0{\hfil/\hfil}} #1 \else
      \rlap{\hbox to \dimen1{\hfil$#1$\hfil}} / \fi}
\ifx\oldchi\undefined \let\oldchi=\chi
  \def\cchi{{\raise 1pt\hbox{$\oldchi$}}} \let\chi=\cchi \fi

\def\frac#1#2{{\textstyle{#1 \over #2}}}

\def\half{\ifinner {\scriptstyle {1 \over 2}}\else {1 \over 2} \fi}

\def\simge{\rlap{\raise 2pt \hbox{$>$}}{\lower 2pt \hbox{$\sim$}}}
\def\simle{\rlap{\raise 2pt \hbox{$<$}}{\lower 2pt \hbox{$\sim$}}}

\def\buildchar#1#2#3{{\null\!\mathop{#1}\limits^{#2}_{#3}\!\null}}

\def\vbig#1#2{{\vbigd@men=#2\divide\vbigd@men by 2%
\hbox{$\left#1\vbox to \vbigd@men{}\right.\n@space$}}}

\null
%
%
%
%

%
\nopagenumbers{\baselineskip=12pt
\line{\hfill PUPT-1286}
\line{\hfill September, 1991}
\ifdoublepage \bjump\bjump\bjump\bjump\else\vfill\fi
\centerline{\capsone QUANTUM SOLITONS IN AFFINE TODA FIELD THEORIES}
\bjump\bjump
\centerline{\tencaps Timothy Hollowood\footnote{$^*$}{Address after
Oct. 1, 1991: Dept. of Theoretical Physics, Oxford, U.K.}}
\sjump
\centerline{\sl Joseph Henry Laboratories, Department of Physics,}
\centerline{\sl Princeton University, Princeton, N.J. 08544}
\vfill
\ifnum\unoduecol=2 \eject\null\vfill\fi
\centerline{\capsone ABSTRACT}
\sjump
\noindent The spectra of $A_r$ affine Toda field theories with
imaginary coupling constant, are investigated. Soliton solutions
are found, which, despite the non-unitary form of the Lagrangian,
have real classical masses and are stable to small perturbations. The quantum
corrections to the soliton masses are determined, to lowest order in $\hbar$.
The solitons have the same spectrum as the fundamental
Toda particles; a feature that is preserved in the quantum theory.
\sjump
\ifnum\unoduecol=2 \vfill\fi
\eject}
\yespagenumbers\pageno=1

\introduction

Amongst the conformal field theories in two dimensions,
there exists a much simpler class of theories,
the so-called `minimal theories', whose spectra consist of a finite number of
representations of the -- possibly extended -- conformal algebra
[\Ref{bpz}]. For example, there are the theories where the conformal
algebra is extended to the $W_{r+1}$-algebras [\Ref{walgebra}].
For each $r$ there exists a whole set of such theories whose central
charges are conveniently parameterized in terms of a constant $\alpha$:
$$
c=r\left\{1+(r+1)(r+2)\left(\alpha+{1\over\alpha}\right)^2\right\}.
\efr
For $c\geq r$ -- so $\alpha$ is real -- the
theories are unitary, but not minimal. For $c<r$ -- so $\alpha$
is purely imaginary -- the situation is more complicated: when $\alpha^2$
is rational the theory is minimal, since
a finite set of representations decouple from the
complete Hilbert space.
Furthermore, when $\alpha^2$ takes the particular rational values
$$
\alpha^2=-{r+1+k\over r+2+k},
\nfr{ratcoup}
where $k$ is a positive integer, the theory is, in addition,
unitary. So, although a generic theory for which $c<r$ is non-unitary,
there exists a set of measure zero unitary theories, when the coupling
constant takes the rational values of \ratcoup.

This paper constitutes a study of the analogue of this phenomenon in the
space of massive two-dimensional theories. In fact,
we shall consider the massive theories which are thought to describe
the $W_{r+1}$-conformal field theories perturbed by the particular
relevant operator which preserves the integrability of the theory
[\Ref{wpert}]. With $\alpha$ as in \ratcoup, the
operator has dimension $(k+1)/(r+k+1)$. A deformation which
preserves integrability is very special, and suggests that the
resulting massive theory may be tractable.

There is evidence to suggest that these integrable theories are
actually affine Toda field theories [\Ref{todapert}]. The case when
$r=1$, which is the sine-Gordon theory, has
been studied in some detail [\Ref{sgpert}], however, the
more theories for more general $r$ have not been the subject of much study
-- one of the reasons being that they are manifestly non-unitary.
We show that although the theories are
non-unitarity, there does exist a sector of soliton
excitations which appears to be unitary. This suggests that the
soliton spectrum is the massive remanent of the primary conformal
fields of the minimal conformal field theory which describes the
ultra-violet scaling limit of the theory.

The Lagrangian density of an affine Toda field theory -- for a
simply-laced Lie algebra -- can be written in a number of ways: for instance
$$
{\cal L}=\frac12\left((\partial_t\phi)^2-(\partial_x\phi)^2\right)-
{m^2\over\gamma^2}\sum_{j=0}^rn_j\left(e^{\gamma\alpha_j\cdot\phi}-
1\right).
\nfr{lag}
The field $\phi(x,t)$ is an
$r={\rm rank}(g)$ vector, where $g$ is a simply-laced Lie algebra,
and consequently
an algebra of type $A$, $D$ or $E$. The inner products are taken
with respect to the Killing form of $g$ restricted to the Cartan
subalgebra. The $\alpha_j$'s, for $j=1,\ldots,r$ are the simple roots
of $g$; $\alpha_0$ is the extended root (minus the highest root of
$g$). The fact that the extended root is included in the sum,
distinguishes the {\it affine\/} theories from the {\it non-affine\/} ones.
The $\alpha_j$'s are linearly dependent:
$$
\sum_{j=0}^rn_j\alpha_j=0,\ \ \ \ n_0=1,
\efr
which serves to define the integers $n_j$; $\gamma$ and $m$ are constants.
The quantum field theory defined by \lag, is then thought to describe
the integrable perturbation of the $W(g)$-algebra conformal field
theory\note{The $W_{r+1}$-algebra theories are those for which $g=A_r$.},
with the identification $\gamma=\sqrt{4\pi}\alpha$.
Notice that the particular value of
$\gamma$ required to reproduce the minimal theories is actually
imaginary. This implies that the Lagrangian
density is no longer real -- except for $g=A_1$, when it is the
sine-Gordon theory. {\it Prima facie\/}, this would imply that the quantum
field theory defined by \lag\ would be non-unitary. To mirror the
situation at the critical point, we propose that when
$\gamma^2/4\pi$ takes the rational values of \ratcoup, the Toda
theory admits a unitary restriction. With this as the ultimate goal we
now pursue the more modest aim of understanding the spectrum of the
Toda theory with imaginary coupling constant $\gamma$.

Before proceeding it is worth briefly reviewing the properties of the
quantum
theory for real values of $\gamma$. The approach that has been adopted
in the literature entails considering \lag\ in the weak coupling
limit, and then comparing conjectured forms for the full $S$-matrix
with the weak coupling perturbation expansion [\Ref{todasm}].
The reason why this strategy is
feasible is because the theory is expected to be integrable also at the
quantum level, and so the $S$-matrix should be {\it factorizable\/}
and also, in this instance, {\it purely elastic\/}
[\Ref{todasm}]. The weak coupling expansion involves expanding \lag\ in
powers of $\gamma$ about the minimum of the potential at $\phi=0$; the first
two terms are
$$
V(\phi)={m^2\over2}\sum_{j=0}^rn_j(\alpha_j\cdot\phi)^2
+{m^2\gamma\over6}\sum_{j=0}^rn_j(\alpha_j\cdot\phi)^3+O(\gamma^2).
\efr
In this limit, the theory describes a set of $r$ particles -- the fundamental
Toda particles -- whose masses
are given by $2m\sqrt\zeta$, where the $\zeta$'s are the
the eigenvalues of the matrix
$$
\sum_{j=0}^rn_j\alpha_j\otimes\alpha_j,
\nfr{massmat}
interacting via a set of three-point couplings defined by the term of
order $\gamma$. The higher order terms describe corrections at various
powers of $\gamma$. The first remarkable fact is that in the one-loop
approximation all the masses are renormalized by an overall constant
[\Ref{todasm}]. (This result is peculiar to the simply-laced theories.)
On the basis of this,
the first conjecture is that the mass ratios are preserved to all orders
in $\gamma$. Furthermore, there are no
additional states, so the quantum theory preserves the classical
spectrum. For example, in the $A_r$ theories the classical masses are
$$
m_a=2m\sin\left({\pi a\over r+1}\right),\ \ \ \ a=1,2,\ldots,r,
\nfr{classmass}
and in the quantum theory the bare mass $m$ is replaced by a
renormalized mass $m_0$. Recently, the
conjectured $S$-matrix of the fundamental particles
has been written in an elegant universal
way using properties of the root system of the Lie algebra [\Ref{dorey}].

If $\gamma$ is imaginary, say $\gamma=i\beta$ for real $\beta$, then
a na\"\i ve analytic continuation of the particle $S$-matrix
would lead one to conclude that the spectrum discussed above cannot be
complete because of the appearance of new poles on the physical strip,
indicating the existence of new particles states. Moreover, some the
the residues of the $S$-matrix, at the poles on the physical-strip, change
sign -- the hallmark of non-unitarity. In fact, we shall
argue that the spectrum of the theory is actually much
richer than this, because with imaginary coupling constant the theory also
admits solitons. The solitons are present at the classical level, and
correspond to static solutions of the equations of motion, in some
frame of reference, with finite energy. It is straightforward to see
why solitons will exist: with imaginary coupling constant the
potential of the theory is periodic $\phi\sim\phi+(2\pi/\beta)\omega$,
where $\omega$ is an element of the weight lattice $\Lambda^\star$ of
$g$. So the field $\phi$ takes values on the maximal torus of $g$.
Solitons correspond to field configurations which wind around the
non-trivial cycles of the maximal torus, as $x$ goes from $-\infty$ to
$\infty$.
Thus the solitons are `kink' solutions, generalizing the situation for the
sine-Gordon theory.  What is not clear --
since the potential is complex -- is that the
classical solitons have a {\it real\/} positive energy, and that they
are stable to small perturbations. Remarkably both the former and latter
are true for the $A_r$ theories, as we shall
show. It turns out that the mass spectrum of solitons is equal to
\classmass, up to an overall factor, although the solitons carry an
additional quantum number which measures the topological charge.
We further show that quantum corrections do not destroy this
structure.

The paper is organized as follows. In \S1, the classical theory is
considered, multiple soliton solutions are constructed and the
classical soliton masses are calculated. In \S2 the quantum
corrections to the soliton masses are calculated. \S3 contains
some comments.

\chapter{The Classical Solitons}

In this section we will construct classical soliton solutions of the
$A_r$ theories described by \lag\ and calculate their masses.

\section{Multiple Soliton Solutions}

Our approach to constructing the soliton solutions follows
the methods developed by R. Hirota for finding explicit expressions for the
multi-soliton solutions of integrable equations [\Ref{hirota}].
There are alternative formalisms
for finding these solutions, most notably the inverse scattering
method, however, for this particular task the Hirota method is far
superior; its elegance being manifested in a
subtle kind of non-linear superposition principle for soliton
solutions, which easily allows one to construct multiple soliton
solutions. The method
is intimately connected with the tau-function approach to integrable
theories, of which an up to date survey may be found in
[\Ref{solbook}].

Before more general theories are treated, it is useful
to recall some properties of the sine-Gordon theory,
which is obtained by setting $r=1$. With the
definition $\psi=\sqrt2\phi$, the equation of motion, which follows
from \lag, is
$$(\partial_t^2-\partial_x^2)\psi+{4m^2\over\beta}\sin\beta\psi=0.$$
It has been known for a long time (see [\Ref{raj}] for example)
that this equation admits a soliton solution of the form
$$\psi={4\over\beta}\tan^{-1}\left(e^{\sigma(x-vt-\xi)}\right),
\nfr{sgsol}
where $\sigma$, $v$ and $\xi$ are constants, and
$\sigma^2(1-v^2)=4m^2$. The anti-soliton corresponds to the negative
of the above. (The direct method was applied to the theory by R.
Hirota in [\Ref{hirotasg}].)

We now turn to the more general theories. The crucial step of the Hirota method
is a change of
variables which converts the equation of motion to an equation of
`Hirota bilinear type' (to use the parlance of the subject). The
appropriate change of variables is not obvious; fortunately
we are guided by the change of variables which is used for the Toda
lattice equations, which are identical to the Toda field theory
equations except that in the former the field only depends on $t$ [\Ref{toda}].
We are led to consider
$$
\phi=-{1\over i\beta}\sum_{j=0}^r\alpha_j\log\tau_j,
\nfr{chvar}
where, for the moment, we are considering a general simply-laced
theory. The new variables are the tau-functions for this theory.
In terms of the $\tau_j$'s, the equations of motion become
$$
\sum_{j=0}^r\alpha_j\left({\ddot\tau_j\tau_j-\dot\tau_j^2-\tau_j
^{\prime\prime}\tau_j+\tau_j^{\prime2}\over\tau_j^2}\right)=m^2
\sum_{j=0}^rn_j\alpha_j\prod_{k=0}^r\tau_k^{-\alpha_j\cdot\alpha_k}.
\nfr{taueq}
(Here, dot denotes a $t$-derivative and prime
an $x$-derivative.) This equation can be decoupled to give the
collection of equations
$$
\ddot\tau_j\tau_j-\dot\tau_j^2-\tau_j^{\prime\prime}\tau_j+\tau_j
^{\prime2}=m^2n_j\left(\prod_{k=0}^r\tau_k^{A_{jk}}-\tau_j^2\right).
\nfr{decoup}
The matrix with elements $A_{jk}$ is the adjacency matrix\note{This
is the matrix which encodes the
connectivity of the Dynkin diagram, such that $A_{jj}=0$, and $A_{jk}=1$
if the $j^{th}$ and $k^{th}$ spot are joined, {\it i.e.\/}
$\alpha_j\cdot\alpha_k=-1$.} for the
Dynkin diagram of the affine algebra $g^{(1)}$.
In decoupling \taueq, the coefficient of
$\tau_j^2$ has been fixed without loss of generality.

When $g=A_r$ the equations are
$$
\ddot\tau_j\tau_j-\dot\tau_j^2-\tau_j^{\prime\prime}\tau_j+\tau_j
^{\prime2}=m^2(\tau_{j-1}\tau_{j+1}-\tau_j^2).
\nfr{antaueq}
The label on $\tau_j$ is to be understood modulo $r+1$, owing to the
periodicity of the extended $A_r$ Dynkin diagram. Those
familiar with the Hirota method will identify \antaueq\ as being of a `bilinear
type', whereas, on the contrary, for the other simply-laced algebras this does
not seem to be the case. We do not consider these other algebras
further, however, we will make some comments about the $D_4$ theory in
\S3.

To find multi-soliton solutions of the equations, the Hirota method
starts by considering a series expansion in some arbitrary parameter
$\epsilon$, which will be set to 1 at the end of the calculation. So
$$
\tau_j=1+\epsilon\tau_j^{(1)}+\epsilon^2\tau_j^{(2)}+\cdots.
\nfr{expan}
A solution describing $N$ solitons results from terminating \expan\ by
setting $\tau_j^{(a)}=0$ for $a>N$. The fact that the series can be
terminated is a crucial property of Hirota bilinear equations.
Furthermore, one of the nice features of this approach is that the
solutions have a very universal form for all integrable theories. The
one-soliton solution has $\tau_j^{(a)}=0$ for $a>1$, and
$\tau_j^{(1)}$ satisfies the linear differential equation
$$
\ddot\tau_j^{(1)}-\tau_j^{(1)\prime\prime}=m^2(\tau_{j+1}^{(1)}+\tau_{j-1}
^{(1)}-2\tau_j^{(1)}),
\efr
and the condition
$$
\tau_{j+1}^{(1)}\tau_{j-1}^{(1)}=\tau_j^{(1)\,2},
\efr
which are solved by
$$
\tau_j^{(1)}=\exp\left\{\sigma(x-vt-\xi)\right\}\delta_j,
\efr
for constants $\sigma$, $v$, $\xi$ and $\delta_j$. The constants
$\delta_j$ have to satisfy the recursion relations
$$
\delta_{j+1}\delta_j^{-1}=\delta_j\delta_{j-1}^{-1},
\efr
and the periodicity requirement $\delta_{j+r+1}=\delta_j$. It is
straightforward to solve for $\delta_j$, yielding
$$\delta_j=\omega^j,
\efr
where $\omega$ is an $(r+1)^{\rm th}$ root of unity. So there are $r$
non-trivial solutions for which $\omega=\omega_a\equiv\exp2\pi i
a/(r+1)$. In addition the constants $\sigma$ and $v$, and
the integer $a\in\{1,2,\ldots,r\}$, have to satisfy
$$
\sigma^2(1-v^2)=4m^2\sin^2\left({\pi a\over n+1}\right)=m_a^2,
\nfr{disp}
where the $m_a$'s are the classical masses of the fundamental particles
defined in \classmass.
\disp\ is just an expression of the fact that the theory is Lorentz
invariant; the faster a soliton goes the narrower it becomes.

The one-soliton solution is therefore
$$
{\overline\phi}_{(a)}(x,t)=-{1\over i\beta}\sum_{j=0}^r\alpha_j
\log\left(1+e^{\sigma(x-vt-\xi)}\omega_a^j\right).
\nfr{onesol}
The solution is in general complex, however, the asymptotic values of
${\overline\phi}_{(a)}$ as $x\rightarrow\pm\infty$ are real.
If we assume that $\sigma>0$, as $x\rightarrow-\infty$
$\phi\rightarrow0$, and as $x\rightarrow\infty$
$$
\phi\rightarrow {2\pi\over\beta}\left(\lambda_a\ {\rm
mod}\,\Lambda\right),
\efr
where $\lambda_a$ is the $a^{\rm th}$ fundamental weight\note{The fundamental
weights are dual to the roots: $\alpha_j\cdot\lambda_k=\delta_{jk}$.}
 and $\Lambda$ is
the root lattice of $g$. We define the topological charge a
configuration to be
$$
t={\beta\over2\pi}\int_{-\infty}^\infty\,dx\,\partial_x\phi=
{\beta\over2\pi}\left(\lim_{x\rightarrow\infty}-\lim_{x\rightarrow-\infty}
\right)\phi(x).
\nfr{topch}
So the one-soliton solution ${\overline\phi}_{(a)}$ has a topological
charge which is equal to the $a^{\rm th}$ fundamental weight, up to a
root. The actual value depends on a choice for the branch-cuts of the
logarithms in \onesol. A careful analysis of the possibilities shows
that the possible topological charges of the one-soliton solution
${\overline\phi}_{(a)}$ fill out the weights of the $a^{\rm th}$
fundamental representation.

When $g=A_1$ the solution can be chosen to real by choosing
$\exp\sigma\xi=-i\exp\sigma\xi^\prime$ with $\xi'$ real, {\it i.e.\/}
$$
\tau_1=\tau_0^\star=1-ie^{\sigma(x-vt-\xi^\prime)},
\efr
reproducing the sine-Gordon soliton solution \sgsol.

Multi-soliton solutions result from terminating \expan\ at a higher power of
$\epsilon$. It might be thought that the complexity of the solution
increases rapidly with the number of solitons; however, this is not
the case due to a subtle form of non-linear superposition principle
which underlies the Hirota method. The two soliton
solution illustrates the situation. The solution depends on a set of
data $\{\sigma_p,v_p,\xi_p,a_p\}$, with $p=1,2$; $v_p$ are the velocities,
$\xi_p$ are the position offsets and $\sigma_p$ are the size parameters of
the solitons. It is useful to define
$$
\Phi_j^{(p)}=\sigma_p(x-v_pt-\xi_p)+\log\left(\omega_{a_p}^j\right),
\efr{
then the form of the two soliton solution, after setting $\epsilon=1$,
is
$$
\tau_j=1+e^{\Phi_j^{(1)}}+e^{\Phi_j^{(2)}}+Ae^{\Phi_j^{(1)}+
\Phi_j^{(2)}}.
\nfr{twosol}
Substituting this ansatz into \antaueq\ fixes the `interaction constant'
$$
A=-{(\sigma_1-\sigma_2)^2-(\sigma_1v_1-\sigma_2v_2)^2-4m^2
\sin^2\left({\pi\over n+1}(a_1-a_2)\right)\over(\sigma_1+\sigma_2)^2-
(\sigma_1v_1+
\sigma_2v_2)^2-4m^2\sin^2\left({\pi\over n+1}(a_1+a_2)\right)}.
\nfr{intconst}
The expression for the multi-soliton solution builds on this elegant
form: for example, the three soliton solution is
$$
\eqalign{\tau_j&=1+e^{\Phi_j^{(1)}}+e^{\Phi_j^{(2)}}+e^{\Phi_j^{(3)}}+
A^{(12)}e^{\Phi_j^{(1)}+\Phi_j^{(2)}}+\cr
&\ \ \ \ A^{(13)}e^{\Phi_j^{(1)}+\Phi_j^{(3)}}+A^{(23)}e^{\Phi_j^{(2)}+
\Phi_j^{(3)}}+A^{(12)}A^{(13)}A^{(23)}e^{\Phi_j^{(1)}+\Phi_j^{(2)}+\Phi_j
^{(3)}}.\cr}
\efr
The general $N$ soliton solution can be written in the compact form
$$
\tau_j=\sum_{\{\mu_p\}=0,1}\exp\left(\sum_{p=1}^N\mu_p\Phi_j^{(p)}
+\sum_{p,q=1}^N\mu_p\mu_q\gamma^{(pq)}\right),
\efr
where $\gamma^{(pq)}=\log A^{(pq)}$. The sum is
over the $2^N$ possibilities for which $\mu_p=0$
or 1, for each $p$.

The multi-soliton solution have a natural physical interpretation.
They represent the histories of a set of solitons which scatter off each other.
To make this more precise consider the two soliton solution. Suppose that
$\xi_1<\xi_2$, $v_1>v_2$, and $\sigma_1$ and $\sigma_2$ are both
positive. Focus on the solution in the vicinity of the first soliton, {\it
i.e.\/} $x\sim v_1t+\xi_1$. Initially, that is in the limit
$t\rightarrow-\infty$, the solution is approximately
$$
\tau_j\simeq 1+e^{\Phi_j^{(1)}},
$$
and finally, {\it i.e.\/} as $t\rightarrow\infty$,
the solution is approximately
$$
\tau_j\simeq
e^{\Phi_j^{(2)}}\left(1+Ae^{\Phi_j^{(1)}}\right).
$$
In both the limits, the solution represents an isolated soliton, the
only difference is that the final `position offset' has been
displaced: $\xi_1\longmapsto \xi_1+\ln A$.
The fact that the velocities and form are unchanged by the
collision is a characteristic feature of soliton scattering in  an integrable
field theory, reflecting the existence of the infinite set of
integrals of motion. So asymptotically the solution represents
two isolated solitons. What is not so apparent is that the topological
charges of the solitons can be altered by the scattering (although the
representations, {\it i.e.\/} $a_1$ and $a_2$, are fixed). To see this
requires a rather detailed analysis, that will not be pursued here,
of the positions of the branch-cuts of the logarithms.

\section{Classical Soliton Masses}

The purpose of this section is to calculate the classical
masses of the solitons
that were constructed in the last section. The result is highly
significant because it
transpires that the ratios of the soliton masses are equal to the
ratios of the masses of the fundamental particles \classmass. The
result is that the one-soliton solution
of \onesol, with a
topological charge $t=\lambda_a\ {\rm mod}\,\Lambda$, has classical mass
$$
M_a^{\rm cl}={4m(r+1)\over\beta^2}\sin\left({\pi a\over
r+1}\right)={2(r+1)\over\beta^2}m_a.
\nfr{solcm}
To find the masses of the soliton solutions, either the energy
$$
M\gamma(v)=\int_{-\infty}^\infty dx\,\left(\frac12\dot\phi^2+
\frac12\phi^{\prime2}-{m^2\over\beta^2}\sum_{j=0}^rn_j\left(e^{i
\beta\alpha_j\cdot\phi}-1\right)\right),
\efr
or the momentum
$$
M\gamma(v)v=-\int_{-\infty}^\infty dx\,\dot\phi\cdot\phi^\prime,
\efr
of a soliton must be calculated. (As usual, in the above
$\gamma(v)=(1-v^2)^{-1/2}$.) The momentum is somewhat easier
to calculate. In doing so it is useful to introduce the weights of the
$(r+1)$-dimensional representation\note{The weights of the $(r+1)$-dimensional
representation are
$e_j$, for $j=1$ to $r+1$, with $\sum_{j=1}^{r+1}e_j=0$; in terms of which
the roots are $\alpha_j=e_j-e_{j+1}$, with $e_0\equiv e_{r+1}$.}
of $A_r$, and to define $\phi_j=e_j\cdot\phi$. The momentum is then
$$
P=-\int_{-\infty}^\infty dx\,\sum_{j=1}^{r+1}\dot\phi_j\phi_j^\prime
=-\sum_{j=1}^{n+1}\int_{\phi_{j-}}^{\phi_{j+}}d\phi_j\dot\phi_j,
\nfr{moment}
where $\phi_{j\pm}$ are the asymptotic values
of $\phi_j(x,t)$ as $x\rightarrow\pm\infty$. Using the explicit
expression for the one-soliton solution $\phi_{(a)}(x,t)$ in \onesol\
one finds
$$
\eqalign{\dot\phi_j&=-{1\over i\beta}{\partial\over\partial t}
\ln\left({\tau_j
\over\tau_{j-1}}\right)\cr &={\sigma v\over i\beta}\left({\delta^je^
\Phi\over1+\delta^je^\Phi}-{\delta^{j-1}e^\Phi\over1+\delta^{j-1}e^\Phi}
\right),\cr}
\nfr{phidot}
where $\delta=\omega_{a}$ and $\Phi=\sigma(x-vt-\xi)$. This
expression can be rewritten in terms of $\phi_j$ using the expressions
$$
{\delta^je^\Phi\over1+\delta^je^\Phi}=-{1\over1-\delta^{-1}}\left(
{1+\delta^{j-1}e^\Phi\over1+\delta^je^\Phi}-1\right)={1\over1-\delta^{-1}
}\left(1-e^{i\beta\phi_j}\right),
\efr
and
$$
{\delta^{j-1}e^\Phi\over1+\delta^{j-1}e^\Phi}=-{1\over1-\delta}\left(
{1+\delta^je^\Phi\over1+\delta^{j-1}e^\Phi}-1\right)={1\over1-\delta
}\left(1-e^{-i\beta\phi_j}\right).
\efr
Using these \phidot\ becomes
$$
\dot\phi_j={\sigma v\over i\beta}\left({1\over1-\delta}
e^{-i\beta\phi_j}-{1\over1-\delta^{-1}}e^{i\beta\phi_j}+{1\over
1-\delta^{-1}}-{1\over1-\delta}\right).
\nfr{phidottwo}
The utility of this expression is that $\dot\phi_j$ has been written
solely in terms of $\phi_j$, and hence the integrals in \moment\ can now be
performed. The constant terms in \phidottwo\ do not contribute because
$\sum_{j=1}^{r+1}\phi_j=0$. Therefore
$$
P=-{\sigma v\beta^2}\sum_{j=1}^{r+1}\left.\left({1\over1-\delta}e^{-i
\beta\phi_j}+{1\over1-\delta^{-1}}e^{i\beta\phi_j}\right)\right\vert_{\phi_
{j-}}^{\phi_{j+}}.
\efr
The evaluation of this expression is straightforward using the
asymptotic expressions for $\phi_j$
$$
e^{i\beta\phi_j}=\cases{\delta^{-1}\ \ \ \ &$x\rightarrow\infty$
\cr 1 &$x\rightarrow-\infty$.\cr}
\efr
The following masses are found
$$
M_a^{\rm cl}={P\over\gamma(v)v}={2\sigma
(r+1)\over\gamma(v)\beta^2}={4m(r+1)\over\beta^2}\sin\left({\pi a\over
n+1}\right),
\efr
where the relation \disp\ was employed to eliminate $\sigma$.

\chapter{Quantum Corrections to the Soliton Masses}

In this section we calculate the first quantum corrections to the
soliton masses. We employ the semi-classical WKB method described in
[\Ref{wkb}] for the kinks, or solitons, of the $\phi^4$ and
sine-Gordon theories -- see also the review in [\Ref{raj}].

The dimensionless expansion parameter in our theory is
$\hbar\beta^2/m^2$, and so the expansion in $\hbar$ coincides with the
weak coupling expansion. For the sine-Gordon theory it is remarkable
that the semi-classical WKB method, which gives the quantum corrections
to the first order in $\hbar\beta^2/m^2$, is exact! The reason why the
WKB method is exact for the sine-Gordon
theory is, presumably, a consequence of its integrability.
It seems plausible that the same is true for the affine Toda theories. In
any case, we compute the mass corrections following exactly the same
method as was used in [\Ref{wkb}] for the $\phi^4$ and sine-Gordon
theories -- the latter being included in our calculation
as the special case for which $r=1$.

{}From now on we set $\hbar=1$. The WKB method of [\Ref{wkb}] is
straightforward to apply in our situation since the one-soliton
solution is, in its rest frame,
time-independent. The idea is to compute the zero-point energy of the
small oscillations around the classical solution. The sum over all the
modes can be done when an infra-red regulator is introduced; this is
most easily achieved by putting the theory in a box. The
zero-point energy of the vacuum must then be subtracted. The resulting
expression is independent of the size of the box, $L$, as
$L\rightarrow\infty$; however, the final expression has an
ultra-violet divergence. This divergence is not a problem of the
soliton solution {\it per se\/},
rather, it is just a manifestation of the fact that the
bare mass needs to be renormalized. This is achieved simply by
normal-ordering the Lagrangian. The extra correction removes the divergence
and the resulting finite residue which remains is then the mass
correction.

To begin with, we must analyse the small perturbations around the
one-soliton solution. If we write $\phi=\overline\phi_{(a)}+
\eta$, where $\overline\phi_{(a)}$ is the static soliton solution in
\onesol -- with $v=0$ and $\sigma=m_a$ --
and expand in powers of $\eta$, to first order in $\eta$ we find
$$
\left(\partial_t^2-\partial_x^2\right)\eta+m^2\sum_{j=0}^re^{i\beta
\alpha_j\cdot\overline\phi_{(a)}}\alpha_j(\alpha_j\cdot\eta)=0.
\nfr{perteq}
For a generic field theory, it is impossible to find the
solutions to the linearized equation of motion around a classical
solution of the equations of motion. However, we are in the fortunate
position of studying an integrable theory for which we can find the
modes explicitly. Asymptotically, as $x\rightarrow\pm\infty$, \perteq\ becomes
$$
\left(\partial_t^2-\partial_x^2\right)\eta+m^2\sum_{j=0}^r\alpha_j(\alpha_j
\cdot\eta)=0.
\efr
This is nothing but the linearized classical equation of motion of the
theory around the `vacuum' field configuration $\phi=0$. Since $\eta$ is an
$r$-dimensional vector there are $r$ independent solutions
of the full equation \perteq, with asymptotic form
$$
\eta_{b}(x,t)\buildchar{\longrightarrow}{ }{x\rightarrow\pm\infty}e^{i(kx+\nu
t+\delta_{\pm})}\sum_{j=0}^r\omega_b^j\alpha_j,\qquad b=1,\ldots,r,
\nfr{asym}
with frequency
$$
\nu^2=k^2+m_b^2,
\efr
which correspond to the fundamental
particles modes of the Toda theory. We now proceed to find the exact
solutions to the
linearized equation \perteq. With \asym\ in mind, we are led to
consider the two-soliton solution in eq. \twosol. The first soliton will be
the static one-soliton solution that we are expanding around, the second
will act as a small perturbation of the first. We take $\sigma_2=ik$ and
$\sigma_2 v_2=-i\nu$ and treat $\exp\sigma_2\xi_2$ as a small parameter,
in which case
$$
\eta_{b}=-{1\over
i\beta}\sum_{j=0}^r\alpha_j{\delta\tau_j\over\tau_j},
$$
where $\tau_j$ corresponds to the one-soliton solution and
$$
\delta\tau_j=e^{\Phi_j^{(2)}}+Ae^{\Phi_j^{(1)}+\Phi_j^{(2)}}.
\efr
The explicit expression for the $b^{\rm th}$ mode is therefore
$$\eta_{b}(k;x,t)=-{1\over
i\beta}\sum_{j=0}^r\alpha_j\left({1+A\omega_{a}^je^{m_a
(x-\xi)}\over1+\omega_{a}^je^{m_a(x-\xi)}}\right)\omega_{b}^j
e^{i(kx+\nu t)},
\nfr{fullsol}
with frequency $\nu^2=k^2+m_b^2$. The interaction parameter $A$ was found
in \intconst, and for the present solution it is
$$A(k)={m_a^2+m_b^2-m_{a-b}^2-2im_a k\over m_a^2+m_b^2-m_{a+b}^2
+2im_ak}.
\nfr{intpar}
It is straightforward to verify that the solutions \fullsol\ then have
the correct asymptotic behaviour as predicted in \asym.

In addition to the solutions found above, there is a zero-mode solution
(having zero frequency) of the form
$$
\eta\propto\left(\partial\overline\phi_{(a)}\over\partial x\right).
\efr
This solution exists because the theory is translationally invariant and
so there exists a whole family of one-soliton
solutions which only differ in the position of the centre-of-mass. For
the WKB method, it
turns out that one can ignore this mode [\Ref{wkb}]; a more
sophisticated time-dependent treatment shows that
its effect is simply to ensure that the quantum soliton behaves as a
relativistic particle, {\it i.e.\/} has energy levels $E=\sqrt{M^2+P^2}$, where
$M$ is the quantum mass of the soliton and $P$ is its momentum.

Before we proceed to calculate the quantum mass of the soliton, it is
worth pointing out that since there are no modes with imaginary
frequency the solitons are classical stable. This is quite surprising
given the fact that the potential of the theory is complex.

In order to calculate the zero-point energy of the modes it is
necessary to introduce some finite boundary conditions, whence the
modes become discrete and it is possible to enumerate them.
The appropriate boundary conditions in this
case are $\eta(x=0)=\eta(x=L)=0$, where the centre-of-mass of the
soliton lies between inside the box, and $L$ is much
larger than the size of the soliton $\sim m_a^{-1}$. The solutions
satisfying these boundary conditions are
$$
\eta_{b}(k_m;x,t)-\eta_{b}(-k_m;x,t),
\efr
with
$$
k_mL+\rho_{b}(k_m)=\pi m,\qquad\quad m\in{\mmmath Z}.
\nfr{periodic}
where $\rho_{b}(k)={\rm Im}(\log\,A(k))$. Notice that
$\rho_b(-k)=-\rho_b(k)$.

The quantum correction to the soliton mass is then obtained by taking
the zero-point energy of the modes around the soliton solution
and subtracting the zero-point energy of modes around the
vacuum, which satisfy \periodic\ with $\rho=0$. That is
$$
\Delta M_a=\frac{1}{2}\sum_{b=1}^r\sum_{m\geq0}\left(\sqrt{k_m^2+m_b^2}-
\sqrt{(k_m+\rho_{b}(k_m)/L)^2+m_b^2}\right),
\efr
where $k_m$ satisfies \periodic. Since $L$ will eventually be taken to
infinity, we can expand in $L^{-1}$, the leading term in the sum being
$$
-{k\rho_{b}(k)\over\sqrt{k^2+m_b^2}}=-\rho_{b}(k){d\epsilon_{b}(k)\over
dk},
\efr
with $\epsilon_{b}(k)=\sqrt{k^2+m_b^2}$. Now, we take
$L\rightarrow\infty$ in which case we can replace the sum over $m$ by
an integral over $k$:
$$
\sum_{m\geq0}=L\left\{\int_{-\infty}^\infty{dk\over2\pi}+O(L^{-1})\right\}.
\efr
Therefore, the mass correction is
$$
\eqalign{\Delta
M_a&=-\sum_{b=1}^r\int_{-\infty}^\infty{dk\over4\pi}\rho_{
b}(k){d\epsilon_{b}(k)\over dk}\cr
&=-{1\over4\pi}\sum_{b=1}^r\left\{\rho_{b}(k)\epsilon_{b}(k)\big
\vert_{-\infty}^\infty+\int_{-\infty}^\infty dk\,\epsilon_{b}(k){
d\rho_{b}(k)\over dk}\right\}.\cr}
\nfr{twoconts}
Now as $k\rightarrow\pm\infty$, $\epsilon_b(k)\rightarrow|k|$ and
$$
\rho_{(b)}(k)\rightarrow{1\over2m_ak}\left(2m_a^2+2m_b^2-m_{a-b}^2
-m_{a+b}^2\right),
\efr
and thus the first contribution from \twoconts\ is
$$\sum_{b=1}^r\rho_{b}(k)\epsilon_{b}(k)\big\vert_{-\infty}
^\infty={1\over
m_a}\sum_{b=1}^r\left(2m_a^2+2m_b^2-m_{a-b}^2-m_{a+b}^2
\right).
\efr
The sum is straightforward to perform, the result being a contribution
$$
-{m_a(r+1)\over2\pi},
\efr
to the mass. As it stands, the second contribution in \twoconts\
is not well defined because the integral has a logarithmic divergence.
The problem is not due to any special nature of the soliton solution,
indeed it is to be expected and is due to the fact that we have not
renormalized the Lagrangian \lag. We now pause to consider
this aspect in more detail.

Mercifully, the renormalization process in a two-dimensional field
theory is straightforward. The divergences can simply be removed by
normal-ordering the Lagrangian. Working to lowest order in $\beta^2$
and introducing an ultra-violet cut-off $\Lambda$
$$
:e^{i\beta\alpha_j\cdot\phi}:=e^{i\beta\alpha_j\cdot\phi}\left(
1+\frac12\beta^2\sum_{a=1}^r{(\zeta_a\cdot\alpha_j)}^2\Delta_a+O(\beta^4)
\right),
\efr
where
$$
\Delta_a=\int_{-\Lambda}^\Lambda{dk\over4\pi}\cdot{1\over\sqrt{k^2
+m_a^2}},
\efr
and $\zeta_a$ is the $a^{\rm th}$ eigenvector of
$\sum_{j=0}^r\alpha_j\otimes\alpha_j$ of eigenvalue $(m_a/m)^2$.
Hence, to lowest order in $\beta^2$, the renormalized Lagrangian is
equal to
$$
L=\frac12\left((\partial_t\phi)^2-(\partial_x\phi)^2\right)
+{1\over\beta^2}\sum_{j=0}^r(m^2+\partial m^2_j)\left(e^{i\beta\alpha
_j\cdot\phi}-1\right),
\efr
with
$$
\partial m^2_j=\frac12(m\beta)^2\sum_{a=1}^r(\zeta_a\cdot\alpha_j)^2\Delta_a,
\efr
and we have added a constant to bring the energy of the vacuum to
zero. The presence of the additional terms clearly changes the energy
of the soliton solution. These changes must be added to the quantum
correction to the soliton mass. It might be thought that we should now
consider the one-soliton solution for the renormalized Lagrangian;
fortunately, this is unnecessary since the correction due to
change in the form of
the solution does not occur at the lowest order in $\beta^2$, precisely
because the soliton solution is an extremum of the action.

Taking into account the counter-term from the renormalization, the
final expression for the mass correction for the $a^{\rm th}$ soliton
is
$$\eqalign{
\Delta M_a=&-{m_a(n+1)\over2\pi}+\sum_{b=1}^r\int_{-\Lambda}^\Lambda
{dk\over4\pi}\epsilon_b(k){d\rho_b(k)\over dk}\cr
 &\qquad\quad-
{1\over\beta^2}
\sum_{j=0}^r\int_{-\infty}^\infty dx\left(e^{i\beta\alpha_j\cdot
\overline\phi_{(a)}}-1\right)\partial m^2_j.\cr}
\nfr{intmass}

We now pause to calculate the integral
$$
\int_{-\infty}^\infty dx\left(e^{i\beta\alpha_j\cdot\overline\phi_{(a)}}
-1\right).
\nfr{theint}
Firstly
$$
e^{i\beta\alpha_j\cdot\overline\phi_{(a)}}=\prod_{k=0}^r\tau_j^{-\alpha_j
\cdot\alpha_k}={\tau_{j-1}\tau_{j+1}\over\tau_j^2},
\efr
where $\tau_j=1+\omega_{a}^je^{m_a (x-\xi)}$ is the static one-soliton
solution. Using \antaueq\ we see that
$$\eqalign{
e^{i\beta\alpha_j\cdot\overline\phi_{(a)}}-1=&{1\over m^2}\cdot{-\tau_j''
\tau_j+\tau_j'^2\over\tau_j^2}\cr
=&-\left({m_a\over m}\right)^2\cdot{\omega_{a}^je^{m_a
(x-\xi)}\over(1+\omega_{a}^j e^{m_a(x-\xi)})^2}.\cr}
\efr
The integral \theint\ may now be calculated explicitly yielding
$$\left.
\left({m_a\over
m}\right)^2\cdot{1\over m_a(1+\omega_{a}^je^{m_a(x-\xi)})}\right\vert
_{-\infty}^\infty=-{m_a\over m^2}.
\efr

The important point about the resulting expression for the integral is
that it is independent of $j$. The contribution to the mass correction
\intmass\ from the renormalization counter-term can now be simplified:
$$
\eqalign{
-{1\over\beta^2}\sum_{j=0}^r\int_{-\infty}^\infty &dx\left(e^{i\beta
\alpha_j\cdot\overline\phi_{(a)}}-1\right)\partial m^2_j\cr
&=\frac12m_a\sum_{b=1}^r\sum_{j=0}^r(\zeta_b\cdot\alpha_j)^2\Delta_b\cr
&=\frac12m_a\sum_{b=1}^r\left({m_b\over m}\right)^2\int_{-\Lambda}^\Lambda
{dk\over4\pi}\cdot{1\over\sqrt{k^2+m_b^2}},\cr}
\efr
where we used the fact that $\zeta_b$ is an eigenvector of
$\sum_{j=0}^r\alpha_j\otimes\alpha_j$ of eigenvalue $(m_b/m)^2$.

The expression for the mass correction is now
$$\eqalign{
\Delta M_a&=-{m_a(r+1)\over2\pi}
+m_a\sum_{b=1}^r\int_{-\Lambda}
^\Lambda{dk\over4\pi}\left\{
\frac12\left({m_b\over
m}\right)^2\cdot{1\over\sqrt{k^2+m_b^2}}\right.\cr
&\left.-\sqrt{k^2+m_b^2}{4(m_a^2+m_b^2-m_{a+b}^2
)\over\left((m_a^2+m_b^2-m_{a+b}^2)^2+4m_a^2k^2\right)}\right\}.\cr}
\efr
Using the fact that
$$
\sum_{b=1}^r(2m_a^2+2m_b^2-m_{a-b}^2-m_{a+b}^2)=2m_a^2(r+1),
\efr
and
$$
\sum_{b=1}^r\left({m_b\over m}\right)^2={\rm
Tr}\left(\sum_{j=0}^r\alpha_j\otimes\alpha_j\right)=\sum_{j=0}^r\alpha_j\cdot
\alpha_j=2(r+1),
\efr
one can easily verify that for large $|k|$ the integrand behaves as
$k^{-2}$ and so, as required, the logarithmic divergence is cancelled by the
renormalization counter-term. We can now let the
ultra-violet cut-off $\Lambda$ tend to infinity.

The remaining contribution is given in terms of a convergent integral:
$$
\Delta M_a=m_a\left\{-{r+1\over2\pi}+I_a\right\},
\efr
where
$$\eqalign{
I_a=\int_{-\infty}^\infty {dk\over4\pi}\sum_{b=1}^r&
\left\{-\sqrt{k^2+m_b^2}\cdot{4(m_a^2+m_b^2-m_{a+b}^2
)\over\left((m_a^2+m_b^2-m_{a+b}^2)^2+4m_a^2k^2\right)}\right.\cr
&\qquad\left.+\frac12\left({m_b\over
m}\right)^2\cdot{1\over\sqrt{k^2+m_b^2}}\right\}.\cr}
\efr
Unfortunately, it is not possible to evaluate the integral $I_a$
analytically; except when $r=1$ in which case the integrand is
zero. Actually, this case is of interest since it corresponds to the
sine-Gordon theory, and we find
$$
\eqalign{M^{\rm q}&=M^{\rm cl}+\Delta M\cr
          &={8m\over\sqrt2\beta^2}-{\sqrt2m\over\pi}.\cr}
\efr
So if we define $M^{\rm cl}(\beta^2)$ to be the classical mass of
the soliton, then to lowest order in the quantum corrections
$$
M^{\rm q}=M^{\rm cl}(\beta'^2),
\efr
with
$$\beta'^2={\beta^2\over1-\beta^2/4\pi},
\nfr{betaprime}
the well-known result for the sine-Gordon quantum soliton mass. Different
quantization approaches show that this result is
exact [\Ref{coleman}].

Returning to the more general theories, we can solve the integral
numerically. For a large number of values of $r$ we have found that
$I_a$ is actually independent of $a$. The first 10 values are
presented in the table\note{Numerical integration was performed using
{\it Mathematica\/}$^{\rm TM}$.}.

\def\thickhrule{\noalign{\hrule height1pt}}

\def\medruledgap{height3pt width1pt&\omit&&\omit&width1pt\cr}
\def\bigruledgap{height4pt width1pt&\omit&&\omit&width1pt\cr}
\midinsert
$$
\vtop{\offinterlineskip
\halign{&\vrule#&\strut\quad#\hfil\quad\cr
\thickhrule
\bigruledgap
width1pt& $r$ && $I$ &width1pt\cr
\bigruledgap
\thickhrule
\medruledgap
width1pt& 1 && 0 &width1pt\cr
width1pt& 2 && -1.814 &width1pt\cr
width1pt& 3 && -1.571 &width1pt\cr
width1pt& 4 && -2.672 &width1pt\cr
width1pt& 5 && -2.721 &width1pt\cr
width1pt& 6 && -3.620 &width1pt\cr
width1pt& 7 && -3.792 &width1pt\cr
width1pt& 8 && -4.593 &width1pt\cr
width1pt& 9 && -5.577 &width1pt\cr
\medruledgap
\thickhrule}}
$$
\sjump
\centerline{Table: the integral $I(r)$}
\endinsert
The important point is not the actual value for $I_a$, but the
fact that it is independent of $a$. This means that to lowest order
the soliton spectrum just receives an overall renormalization
and the masses are still proportional to the masses $m_a$. Our final
results is
$$
M_a^{\rm q}=m_a\left({2(r+1)\over\beta'^2}+I\right)+O(\beta^2),
\nfr{themass}
where $\beta'$ is defined in \betaprime. $I$ is the constant which
only depends on $r$.

\chapter{Discussion}

We have demonstrated that Toda theories with imaginary
coupling constant admit soliton solutions, whose topological charges
lie in the set of weights of the fundamental representations of the
algebra. Surprisingly the solitons have real classical masses and are
stable to perturbation. The gross spectrum of the solitons is the same
as that of the fundamental Toda particles. This feature is not spoiled
by the quantum theory, to lowest order in $\hbar$, and with the sine-Gordon
example in mind, we can speculate that, as a result of the integrability of
the theory, the masses in \themass\ are also exact.

In the sine-Gordon theory, the soliton and anti-soliton can form bound
states -- the so-called breathers. The masses of the breathers can be
found using more sophisticated time-dependent WKB methods [\Ref{wkb}].
Remarkably the ground state of the breather spectrum is identified as
the fundamental particle of the sine-Gordon theory. It is clearly of
interest to repeat the calculation of ref. [\Ref{wkb}] and establish
whether the more general Toda theories admit breather states. One
would also like to determine the soliton $S$-matrix. Since the theory
is integrable we expect this $S$-matrix to be factorizable, however,
the solitons carry topological charge and consequently we do not expect the
$S$-matrix to be elastic, {\it i.e.\/} we expect non-trivial
scattering in  the space of topological charge.
It appears that the correct
$S$-matrix has the form of a product of the {\it minimal\/}
$S$-matrix, which determines the pole structure, and a {\it quantum
group\/} $R$-matrix which acts in the space of the topological
charge; details of this will be presented elsewhere. When the
$S$-matrix has been determined, it should then be possible to
determine out whether
the solitons can be decoupled from the non-unitary sector of the theory.

We have only considered the $A_r$ affine Toda field theories and it
would be
interesting to know whether the construction of the solitons can be
extended to the other simply-laced theories. Initial considerations
for the affine theory based on $D_4$ indicate that soliton solutions
do exist, and are also associated to the fundamental representations of the
algebra. Furthermore, the classical masses of the solitons are consistent
with the conjecture
$$M^{\rm cl}_a={2h\over\beta^2}m_a,
\efr
where $h$ is the Coxeter number of the algebra, and the $m_a$'s are
the masses of the fundamental particles of the theory; that is $2m\sqrt\zeta$,
where $\zeta$ is one of the $r$ eigenvalues of \massmat.

Soliton-type solutions of the $A_2$ affine Toda field theory are also
considered in [\Ref{japsol}]; however, they are not the same as the
solutions considered here, since they do not satisfy the classical
equations of motion.

I would like to thank Mark de Groot for some helpful comments on the
manuscript. This work was supported by NSF PHY90-21984.

\references

\beginref
\Rref{wpert}{V.A. Fateev and A.B. Zamolodchikov, `{\sl Conformal field
theory and purely elastic $S$-matrices\/}', Int. J. Mod. Phys. {\bf A5}
(1990) 1025}
\Rref{todapert}{T. Eguchi and S-K. Yang, `{\sl Deformations of
conformal field theories and soliton equations\/}', Phys. Lett. {\bf
224B} (1989) 373\newline
T.J. Hollowood and P. Mansfield, `{\sl Rational conformal field
theories at, and away from, criticality, as Toda theories\/}',
Phys. Lett. {\bf 226B} (1989) 73}
\Rref{todasm}{A.E. Arinstein, V.A. Fateev and A.B. Zamolodchikov, `{\sl
Quantum $S$-matrix of the $1+1$-dimensional Toda chain\/}', Phys. Lett.
{\bf87B}
 (1979) 389\newline
H.W. Braden, E. Corrigan, P.E. Dorey and R. Sasaki, `{\sl Affine Toda
field theory and exact $S$-matrices\/}',
Nucl. Phys. {\bf B338} (1990) 689; `{\sl Multiple poles and other features
of affine Toda field theory\/}', Nucl. Phys. {\bf B356} (1991) 469}
\Rref{hirota}{R. Hirota, `{\sl Direct methods in soliton theory\/}',
in `{\sl Soliton\/}' page 157: ed. R.K. Bullough and P.S. Caudrey
(1980)}
\Rref{hirotasg}{R. Hirota, `{\sl Exact solution of the sine-Gordon
equation for multiple collisions of solitons}', J. Phys. Soc. Japan {\bf33}
(1972) 1459}
\Rref{solbook}{`{\sl Soliton theory: a survey of results\/}', ed. A.P.
Fordy, Manchester University Press (1990)}
\Rref{wkb}{R.F. Dashen, B. Hasslacher and A. Neveu, `{\sl Particle
spectrum in model field theories from semiclassical functional
integral techniques\/}', Phys. Rev. {\bf D11} (1975) 3424; `{\sl
Nonperturbative methods and extended-hadron models in field theory.
II. Two-dimensional models and extended hadrons\/}', Phys. Rev. {\bf
D10} (1974) 4130}
\Rref{raj}{R. Rajaraman, `{\sl Some non-perturbative methods in quantum
field theory\/}', Phys. Rep. {\bf 21C} (1975) 227}
\Rref{coleman}{S. Coleman, `{\sl The quantum sine-Gordon equation as
the massive Thirring model\/}', Phys. Rev. {\bf D11} (1975)
2088}
\Rref{dorey}{P. Dorey, `{\sl Root systems and purely elastic
$S$-matrices\/}', Nucl. Phys. {\bf B358} (1991) 654}
\Rref{toda}{M. Todd, `{\sl Theory of non-linear lattices\/}', Springer
series in solid-state sciences 20}
\Rref{walgebra}{V.A. Fateev and A.B. Zamolodchikov, `{\sl Conformal
quantum field theory models in two-dimensions having $Z(3)$
symmetry\/}', Nucl. Phys. {\bf B280}(FS18) (1987) 644\newline
V.A. Fateev and S.L. Lukyakhov, `{\sl The models of two-dimensional
conformal quantum field theory with $Z(N)$ symmetry\/}', Int. J. Mod.
Phys. {\bf A3} (1988) 507}
\Rref{bpz}{A.A. Belavin, A.M. Polyakov and A.B. Zamolodchikov, `{\sl
Infinite conformal symmetry in two-dimensional quantum field
theory\/}', Nucl. Phys. {\bf B241} (1984) 333}
\Rref{sgpert}{A. LeClair, `{\sl Restricted sine-Gordon theory and the
minimal conformal series\/}', Phys. Lett. {\bf230B} (1989) 103\newline
D. Bernard and A. LeClair, `{\sl The fractional supersymmetric sine-Gordon
models\/}', Phys. Lett. {\bf247B} (1990) 309\newline
F.A. Smirnov, `{\sl Reductions of the sine-Gordon model as perturbations of
conformal field theory\/}', Nucl. Phys. {\bf B337} (1990) 156;
`{\sl The perturbed $c<1$ conformal field theories as reductions of the
sine-Gordon model\/}', Int. J. Mod. Phys. {\bf A4} (1989) 4213\newline
T. Eguchi and S-K. Yang, `{\sl Sine-Gordon theory at rational values
of the coupling constant and minimal conformal models\/}', Phys.
Lett. {\bf235B} (1990) 282}
\Rref{japsol}{T. Nakatsu, `{\sl Quantum group approach to affine Toda
field theory\/}', Nucl. Phys. {\bf B356} (1991) 499}
\endref
\ciao
